\documentclass[twocolumn]{aastex63}

\usepackage{booktabs}
\usepackage{array}
\usepackage{multirow}
\usepackage{nameref}
\usepackage[T1]{fontenc}
\usepackage{breqn}

\submitjournal{ApJ}

\newcommand{\p}[1]{{\color{magenta}{#1}}}

\newcommand{\ie}{{i.e.},}

\shortauthors{To et al.}
\graphicspath{{./}{figures/}}

\begin{document}

\title{Understanding the Relationship between Solar Coronal Abundances and F10.7~cm Radio Emission}

\author[0000-0003-0774-9084]{Andy S.H. To}
\affiliation{University College London, Mullard Space Science Laboratory, Holmbury St. Mary, Dorking, Surrey, RH5 6NT, UK}

\author[0000-0001-7927-9291]{Alexander W. James}
\affiliation{University College London, Mullard Space Science Laboratory, Holmbury St. Mary, Dorking, Surrey, RH5 6NT, UK}
\affiliation{European Space Agency (ESA), European Space Astronomy Centre (ESAC), Camino Bajo del Castillo, s/n, 28692 Villanueva de
la Ca\~nada, Madrid}

\author[0000-0002-0713-0604]{T. S. Bastian}
\affiliation{National Radio Astronomy Observatory (NRAO), 520 Edgemont Road, Charlottesville, VA 22903, USA}

\author[0000-0002-2943-5978]{Lidia van Driel-Gesztelyi}
\affiliation{University College London, Mullard Space Science Laboratory, Holmbury St. Mary, Dorking, Surrey, RH5 6NT, UK}
\affiliation{LESIA, Observatoire de Paris, Universit\'{e} PSL, CNRS, Sorbonne Universit\'{e}, Univ. Paris Diderot, Sorbonne Paris Cit\'{e}, 5 place Jules Janssen, 92195 Meudon, France}
\affiliation{Konkoly Observatory, Research Centre for Astronomy and Earth Sciences, Hungarian Academy of Sciences, Konkoly Thege \'{u}t 15-17., H-1121, Budapest,
Hungary}

\author[0000-0003-3137-0277]{David M. Long}
\affiliation{Astrophysics Research Centre, School of Mathematics and Physics, Queen’s University Belfast, University Road, Belfast, BT7 1NN, Northern Ireland, UK}
\affiliation{University College London, Mullard Space Science Laboratory, Holmbury St. Mary, Dorking, Surrey, RH5 6NT, UK}

\author[0000-0002-0665-2355]{Deborah Baker}
\affiliation{University College London, Mullard Space Science Laboratory, Holmbury St. Mary, Dorking, Surrey, RH5 6NT, UK}

\author[0000-0002-2189-9313]{David H. Brooks}
\affiliation{College of Science, George Mason University, 4400 University Drive, Fairfax, VA 22030, USA}

\author{Samantha Lomuscio}
\affiliation{National Radio Astronomy Observatory (NRAO), 520 Edgemont Road, Charlottesville, VA 22903, USA}

\author[0000-0002-1365-1908]{David Stansby}
\affiliation{University College London, Mullard Space Science Laboratory, Holmbury St. Mary, Dorking, Surrey, RH5 6NT, UK}

\author[0000-0001-7809-0067]{Gherardo Valori}
\affiliation{Max-Planck-Institut f\"ur Sonnensystemforschung, G\"ottingen, Germany}


\begin{abstract}

Sun-as-a-star coronal plasma composition, derived from full-Sun spectra, and the F10.7 radio flux (2.8 GHz) have been shown to be highly correlated (r = 0.88) during solar cycle 24. However, this correlation becomes nonlinear during increased solar magnetic activity. Here, we use co-temporal, high spatial resolution, multi-wavelength images of the Sun to investigate the underlying causes of the non-linearity between coronal composition (FIP bias) and F10.7 solar index correlation. Using the Karl G. Jansky Very Large Array (JVLA), Hinode/EIS (EUV Imaging Spectrometer), and the Solar Dynamic Observatory (SDO), we observed a small active region, AR~12759, throughout the solar atmosphere from the photosphere to the corona. Results of this study show that the magnetic field strength (flux density) in active regions plays an important role in the variability of coronal abundances, and it is likely the main contributing factor to this non-linearity during increased solar activity. Coronal abundances above cool sunspots are lower than in dispersed magnetic plage regions. Strong magnetic concentrations are associated with stronger F10.7~cm gyroresonance emission. Considering that as the solar cycle moves from minimum to maximum, the size of sunspots and their field strength increase with gyroresonance component, the distinctly different tendencies of radio emission and coronal abundances in the vicinity of sunspots is the likely cause of saturation of Sun-as-a-star coronal abundances during solar maximum, while the F10.7 index remains well correlated with the sunspot number and other magnetic field proxies.


\end{abstract}



\section{Introduction} \label{intro}

F10.7~cm radio flux index is one of the most widely used solar indices to characterize the solar activity. Daily measurements of the Sun-as-a-star F10.7~cm flux stretches back to 1947. Cycle to cycle F10.7~cm observations show that the maximum flux could vary by a factor of 2--3~\citep{Floyd2005JASTP..67....3F,Tapping2013SpWea..11..394T}. On the other hand, recent observations of the solar wind have also shown a cyclic behaviour of elemental abundances~\citep{Kasper2007ApJ...660..901K,McIntosh2011ApJ...740L..23M,Lepri2013ApJ...768...94L}. \citet{Brooks2017Aug} observed a correlation between solar coronal abundances and the F10.7 cm radio flux, implying that coronal abundances change with the solar cycle phase. This in turn suggests that coronal abundances are influenced by magnetic activity and the coronal heating process, with significant implications also for solar-like stars. These stars may also show cyclic effects, and the chemical composition of their coronae likely depends on magnetic activity rather than just the fixed properties of the star. However, a saturation of FIP bias is often observed in high activity stars~\citep{Wood2010Jun, Laming2015Sep,Seli2022A&A...659A...3S}, with a typical FIP bias of $\sim1$ or lower. In fact, this saturation is also observed in~\citet{Brooks2017Aug}. During high to extreme solar activity, the correlation between F10.7 and FIP bias becomes non-linear, Sun-as-a-star FIP bias appears to be saturated, while F10.7~cm flux continues to go up with the solar activity. The reason behind FIP bias saturation of both the Sun and highly active stars remains poorly understood. Investigating and understanding the root cause of the F10.7-FIP bias non-linearity, and be able to account for them provide invaluable insight into the solar activity, stellar magnetic fields and both the solar and stellar coronal heating.

The first step to quantify the non-linearity, is to understand elemental abundances variations. Elemental abundances have long been used as an indicator for the physical processes throughout astrophysics. The benchmark reference for all cosmic applications is the solar chemical composition. Understanding the spatial and temporal variations in the composition of the solar corona provides an insight into different physical processes of the Sun, including reconnections in the corona, how matter and energy flow from the chromosphere, where the plasma is separated according to chemical populations (\ie~fractionated), and out into the heliosphere. The method to study and quantify solar and stellar elemental fractionation is to use the first ionization potential (FIP) of elements in the solar atmosphere. High-FIP elements (\ie~FIP $>$10 eV) maintain their photospheric abundances in the corona, whereas low-FIP elements can have enhanced abundances up to a factor of 4$^{+}$ (\ie~FIP bias).  
This is the well-known FIP effect. Conversely, the inverse FIP (IFIP) effect refers to the relative enhancement of high-FIP or relative depletion of low-FIP elements in solar and stellar coronae. 

The level of enhancement of the low-FIP elements in the Sun's atmosphere is far from uniform. FIP bias values depend on factors such as an active region's age, evolutionary stage and the surrounding of active regions. In open-field coronal holes, FIP bias remains unaltered, maintaining the photospheric value of around 1~\citep{Feldman1998Oct,Brooks2011,Baker2013Nov}. Quiet-Sun regions typically have FIP bias in the range of 1.5--2~\citep{Warren1999Dec,Baker2013Nov,Ko2016Aug}, with the highest FIP bias of 3--4 observed in specific locations in solar active regions~\citep{Baker2013Nov,Baker2015,Baker2018,Baker2021Jan,DelZanna2014May,To2021Apr, Mihailescu2022May}. When an active region begins emerging, it is still dominated by photospheric plasma, and it takes from hours to days for the coronal loops to reach peak elemental fractionation values. As the active region begins to decay, the FIP bias slowly returns to that of the surrounding coronal structure~\citep{Baker2018, Ko2016Aug}. 

The temporal variation of composition could also extend beyond hours and days to solar-cycle time-scales of many years. This was indeed shown by~\citet{Brooks2017Aug}, who used data from the EUV Variability Experiment (EVE; \citealp{Woods2012Jan}) on the Solar Dynamics Observatory (SDO; \citealp{Pesnell:2012}) to determine daily Sun-as-a-star FIP bias values from solar minimum to solar maximum during cycle 24. It was demonstrated that the FIP bias derived from full-Sun spectra is highly correlated (r = 0.88) with the F10.7~cm radio flux, a solar activity proxy, during a four-year interval (2010--2014; \citet{Brooks2017Aug} Supplementary Figure 2). However, the relationship between coronal elemental composition and the F10.7~cm radio flux appears to become nonlinear for the period mid 2011/early 2012, and mid 2013/early 2014, when the Sun approached its maximum activity. The FIP bias did not grow in tandem with the F10.7 radio flux, but instead appeared to saturate.

The second step to understand this non-linearity is to examine the emission mechanisms of the F10.7~cm radio flux. Similar to FIP bias values, F10.7~cm emission varies spatially, depending on different solar structures. There are two contributions to the observed radio emission: thermal bremsstrahlung and gyroresonance emission. Various studies have considered the source of these two emission components, with similar suggestions that the bremsstrahlung emission originates from the plage regions~\citep{Felli1981Jul,Tapping1990Jun,Tapping2003Aug}, while suggesting gyroresonance emission comes from the strong magnetic fields in active regions~\citep{Schmahl1995Oct,Schmahl1998,deWit2014,Schonfeld2015Jul}. \citet{Henney2012Feb} analysed the correlation between the photospheric magnetic field and the F10.7 flux. They characterised magnetic “plage” regions as areas with local field strengths of 25–150 G, and the active region component as originating from field with a strength >150 G, and could predict the bremsstrahlung component of the F10.7~cm emission well, which also correlates well with the solar EUV flux \citep{Schonfeld2015Jul, Schonfeld2017Aug, Schonfeld2019Oct}. However, spatially resolved maps linking F10.7 radio flux to coronal composition have never been investigated. In this paper, we present Hinode/EUV imaging spectrometer (EIS) observations of AR~12759 taken on 2020 April 3 and 7 to investigate the contribution of the F10.7 radio flux (2.8~GHz) to elemental fractionation. As previously noted, the correlation between F10.7 flux and coronal abundances has been observed to change under different solar activity conditions~\citep{Brooks2017Aug}. For the first time, these EIS observations are compared with the Solar Dynamics Observatory EUV, magnetic field data and the spatially resolved Stokes~I and Stokes~V maps of the F10.7 flux observed by the Karl G. Jansky Very Large Array~(JVLA; \citealp{Perley2011Aug}). The observations are presented in Section~\ref{obs}, with results and discussion in Section~\ref{results} and Section~\ref{discussion}, respectively. Conclusions are then presented in Section~\ref{conclusion}.

\section{Observations and Data Analysis}\label{obs}
\begin{figure*}[h!]
    \centering    \includegraphics[width=\textwidth]{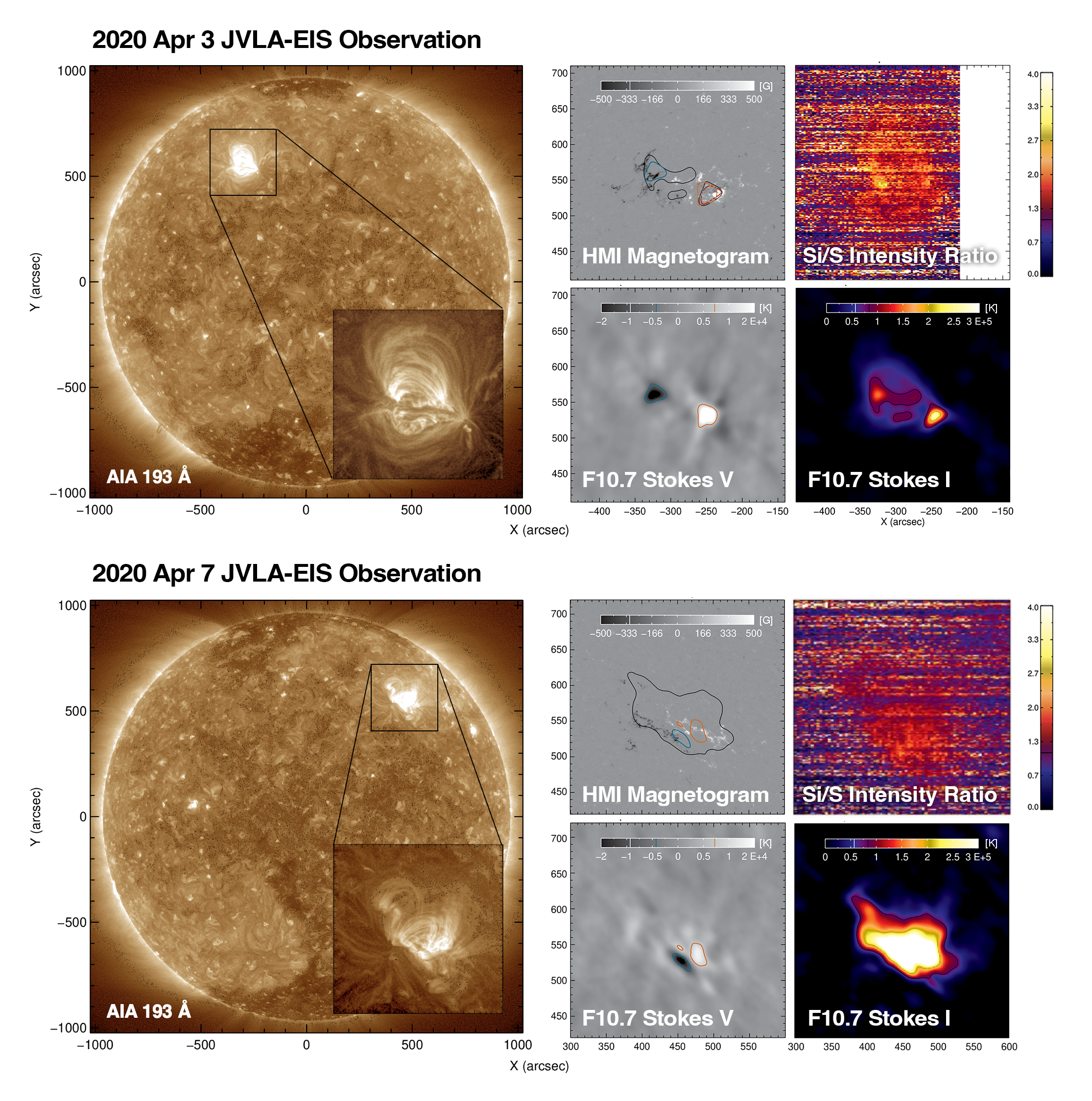}
    \caption{A small bipolar active region, AR~12759, observed during the JVLA/20A-047 observing campaign on April 3 and 7. Left to right, top to bottom: i) AIA 193~\AA; HMI magnetogram; \ion{Si}{10}~258.38~\AA/\ion{S}{10}~264.23~\AA\ intensity ratio map;  F10.7 radio flux map~(Stokes~V); F10.7 radio flux map~(Stokes I). Four F10.7~cm regions are used in this paper to calculate the FIP bias: 1) Stokes~I region with a brightness temperature $\mathrm{>80,000~K}$ subtracted by strong Stokes~V emissions (black contour); 2) Negative Stokes~V regions with brightness temperature $\mathrm{<-10,000~K}$ (blue contour); 3) Positive  Stokes~V region with brightness temperature $\mathrm{>10,000~K}$ (orange contour); and 4) the estimated gyroresonance region on April 3 defined using ((Stokes I-modelled free-free)/Stokes~I)~(red dashed contour; Section~\ref{gyro_region}). It can bee seen that the negative Stokes~V region is associated with the following polarity, whereas the positive Stokes~V and gyroresonance regions are associated with the leading polarity. Si/S intensity ratio maps shown here are for demonstration purposes. In our analysis, we calculated and used the spatially averaged FIP bias. This significantly improves the signal-to-noise ratio.}
    \label{fig:result}
\end{figure*}

\begin{centering}
\begin{table}[]
\resizebox{\columnwidth}{!}{\begin{tabular}{@{}lll@{}}
\multicolumn{3}{c}{} \\ 
\toprule
Study Number & \multicolumn{2}{l}{569} \\
Raster Acronym & \multicolumn{2}{l}{HPW021VEL260x512v2} \\
\multicolumn{1}{c}{} & \multicolumn{2}{l}{\ion{Fe}{8} 185.213~\AA, \ion{Fe}{8} 186.601~\AA} \\
\multicolumn{1}{c}{} & \multicolumn{2}{l}{\ion{Fe}{9} 188.497\AA, \ion{Fe}{9} 197.862~\AA} \\
\multicolumn{1}{c}{} & \multicolumn{2}{l}{\ion{Fe}{10} 184.536~\AA, \ion{Fe}{11} 188.216~\AA} \\
                     & \multicolumn{2}{l}{\ion{Fe}{12} 192.394~\AA, \ion{Fe}{12} 195.119~\AA} \\
Emission Lines       & \multicolumn{2}{l}{\ion{Fe}{13} 202.044~\AA, \ion{Fe}{13} 203.826~\AA}  \\
                     & \multicolumn{2}{l}{\ion{Fe}{14} 264.787~\AA, \ion{Fe}{14} 270.519~\AA} \\
                     & \multicolumn{2}{l}{\ion{Fe}{15} 284.16~\AA, \ion{Fe}{16} 262.984~\AA} \\
                     & \multicolumn{2}{l}{\ion{Fe}{17} 254.870~\AA} \\
                     & \multicolumn{2}{l}{\ion{Si}{10} 258.38~\AA, \ion{S}{10} 264.23~\AA} \\
                     & \multicolumn{2}{l}{\ion{Ca}{14} 193.87~\AA, \ion{Ar}{14} 194.40~\AA} \\
Field of View        & \multicolumn{2}{l}{260\arcsec\ $\times$ 512\arcsec} \\
Rastering            & \multicolumn{2}{l}{2\arcsec\ slit, 87 positions, 3\arcsec\ coarse steps} \\
Exposure Time        & \multicolumn{2}{l}{60~s} \\
Total Raster Time    & \multicolumn{2}{l}{1 hours} \\
Reference Spectral Window   & \multicolumn{2}{l}{\ion{Fe}{12} 195.12~\AA} \\  
\midrule
Study Number    & 544 \\
Raster Acronym &  AbundRaster\_v3 \\
\multicolumn{1}{c}{} & \multicolumn{2}{l}{\ion{Fe}{8} 185.213~\AA, \ion{Fe}{8} 186.601~\AA} \\
\multicolumn{1}{c}{} & \multicolumn{2}{l}{\ion{Fe}{9} 188.497\AA, \ion{Fe}{9} 197.862~\AA} \\
\multicolumn{1}{c}{} & \multicolumn{2}{l}{\ion{Fe}{10} 184.536~\AA, \ion{Fe}{11} 188.216~\AA} \\
                     & \multicolumn{2}{l}{\ion{Fe}{12} 192.394~\AA, \ion{Fe}{12} 195.119~\AA} \\
Emission Lines       & \multicolumn{2}{l}{\ion{Fe}{13} 202.044~\AA, \ion{Fe}{13} 203.826~\AA}  \\
                     & \multicolumn{2}{l}{\ion{Fe}{14} 264.787~\AA, \ion{Fe}{14} 270.519~\AA} \\
                     & \multicolumn{2}{l}{\ion{Fe}{15} 284.16~\AA, \ion{Fe}{16} 262.984~\AA} \\
                     & \multicolumn{2}{l}{\ion{Si}{10} 258.38~\AA, \ion{S}{10} 264.23~\AA} \\
                     & \multicolumn{2}{l}{\ion{Ca}{14} 193.87~\AA, \ion{Ar}{14} 194.40~\AA} \\
Field of View     & 492\arcsec\ $\times$ 512\arcsec \\
Rastering         & 2\arcsec\ slit, 123 positions, 4\arcsec\ coarse steps \\
Exposure Time     & 30~s \\
Total Raster Time & 3~hours \\
Reference Spectral Window & \ion{Fe}{12} 195.12~\AA \\
\bottomrule

\end{tabular}}
\caption{Hinode/EIS study details used in this this work.}
\label{table:study_details}

\end{table}
\end{centering}

AR~12759 was a small and simple bipolar active region that was visible on the northern hemisphere of the Sun from 2020 March 30 to April 10 (as shown in the Atmospheric Imaging Assembly~(AIA; \citealp{Lemen2012}) 193~\AA\ and the Helioseismic and Magnetic Imager~(HMI; \citealp{Scherrer2012}) magnetogram of Figure~\ref{fig:result}). The active region was in its early decay phase when it rotated onto the disk, with a positive polarity leading spot containing a light bridge, and
pore-like transient spots in the following (negative polarity) region were present until April 4. An ephemeral region emerged in the AR's trailing part from about 21:00~UT on the 3rd forming pores, which also disappeared on the 4th. No more spots were seen in the AR after that. Two sets of observations were obtained on 2020 April 3 and 7 during a joint observation campaign by the JVLA and Hinode/EIS (EUV Imaging Spectrometer).

\subsection{Coronal EUV Observation and Alignment}
Details of the two EIS observations made at 13:42~UT on 2020 April 3 and 16:01~UT on 2020 April 7 can be found in Table~\ref{table:study_details}. In this study, we use the \ion{Si}{10}~258.38~\AA/\ion{S}{10}~264.23~\AA\ intensity ratio to examine the spatially averaged changes in the coronal ($\sim$1.25-1.5~MK) FIP bias in a few locations (Blue, orange, black and red contours of Figure~\ref{fig:result}). To minimise effects caused by temperature and density variations, 16 consecutive Fe lines from \ion{Fe}{8}--\ion{Fe}{16} were used in the calculation of the differential emission measure~(DEM). We used the Markov Chain Motel Carlo (MCMC) algorithm distributed with the PINTofALE spectroscopy package~\citep{Kashyap1998ApJ...503..450K, Kashyap2000BASI...28..475K}, and contribution functions taken from the CHIANTI Atomic Database, Version 9.0~\citep{Dere1997Oct,Dere2019ApJS..241...22D}. We also used the photospheric abundances of \citet{Grevesse2007Jun}, assuming the density calculated through the fitted \ion{Fe}{13}~202.04~\AA/203.83~\AA\ intensity ratio. As both Fe and Si are low-FIP elements, we scaled the emission measure to reproduce the observed intensity of \ion{Si}{10}~258.38~\AA. The Si/S FIP bias is then the ratio of the predicted to observed intensity of the \ion{S}{10}~264.23~\AA\ emission line. A more detailed description of the procedures to calculate a coronal composition map can be found in~\citet{Brooks2011}. This method minimises the effects of temperature and density when compared to only taking the Si/S intensity ratio.

One source of error is misalignment of the different instruments. As our results compare observations from Hinode/EIS, JVLA and SDO/AIA that are formed at drastically different solar altitudes, several steps were taken to minimise the instrumental offset between the three instruments. First, the AIA coordinate system was used as our base coordinate system. Second, as the active region was stable throughout the EIS raster duration, the \ion{Fe}{12}~195.12~\AA\ intensity maps observed by EIS were aligned with an AIA 193~\AA\ image taken at the beginning raster time. Lastly, to align JVLA to EIS and AIA, we followed the approach of \citet{Schonfeld2015Jul} to estimate the free-free component of the F10.7 flux using AIA DEM~\citep{Hannah2012Mar}. This allowed us to visualise F10.7 flux data in AIA coordinates. F10.7~cm bremsstrahlung emission taken by JVLA was then aligned to this predicted emission. As a final check, the coronal magnetic field of AR 12759 was modelled by extrapolating radial field magnetograms that were taken by HMI on April 3 and April 7 (see Section 2.3). F10.7~cm gyroresonance observations at 2.8~GHz originates from thin iso-gauss layers with constant magnetic field strengths of B=500~G (2nd harmonic) and 333~G (3rd harmonic), respectively~\citep{White1997Aug}. To investigate the sources of F10.7~cm emission, we modelled the coronal magnetic field of AR 12759 using a linear force-free field extrapolation and visualised isosurfaces at 333~G and 500~G in the coronal volume. These dome-like isosurfaces were compared to VLA Stokes~V observations, enabling us to estimate an emission height for the polarised emissions, and thus the correction required to account for the line of sight optical shifting effect. Since the spatial resolution of the F10.7~cm emission is low compared to the EUV, a very small misalignment should not affect the result.

After the alignment between the three instruments had been confirmed, we separate the F10.7~cm contribution into three parts: total intensity (Stokes~I; brightness temperature $\mathrm{>80,000~K}$), leading polarised data (positive Stokes~V; brightness temperature $\mathrm{>10,000~K}$) and following  polarised data (negative Stokes~V; brightness temperature $\mathrm{<-10,000~K}$). The spatially averaged FIP bias of these sub-regions could then be calculated. 

\begin{figure*}[!ht]
    \centering
    \includegraphics[width=0.48\textwidth]{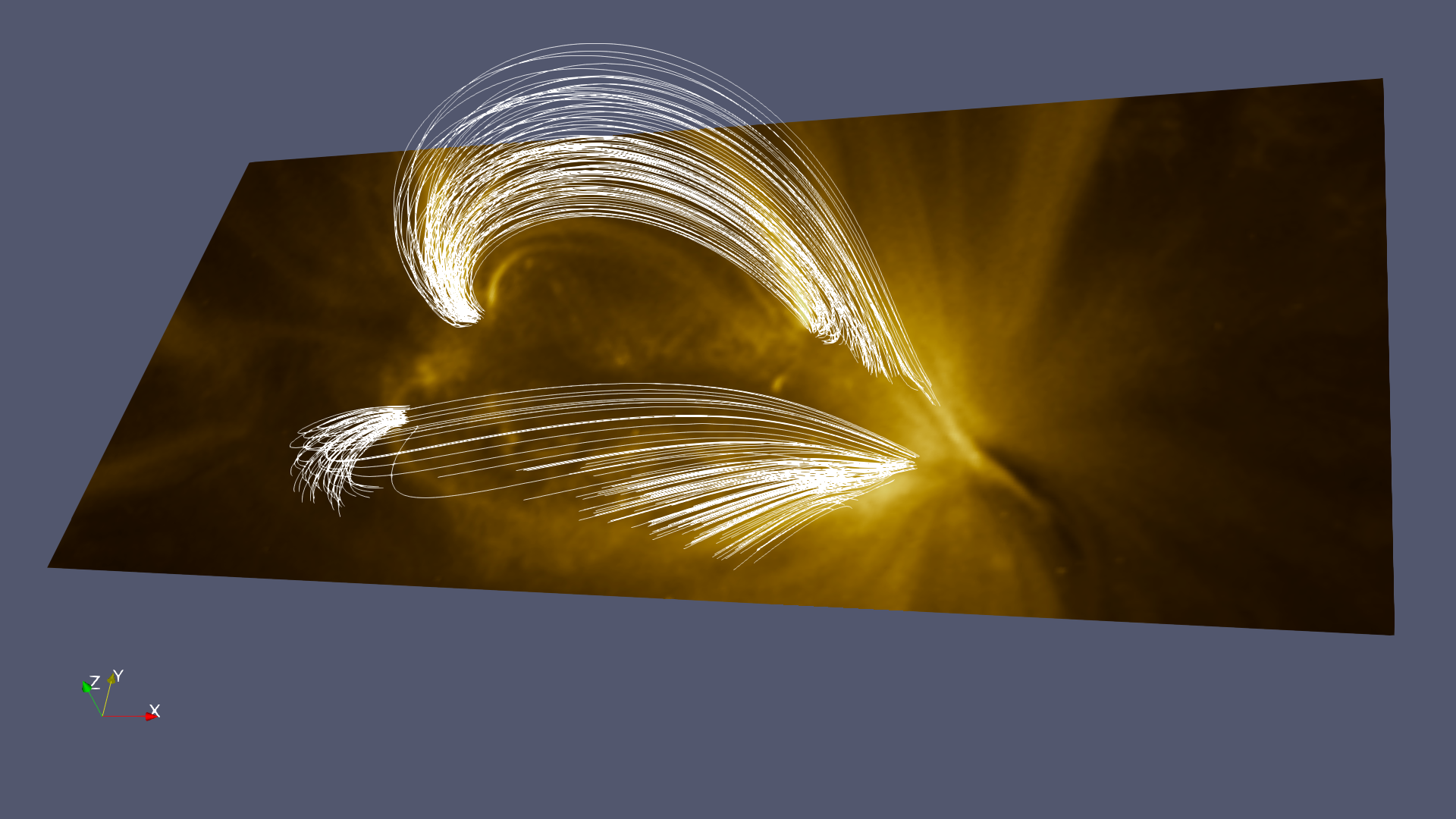}
    \includegraphics[width=0.48\textwidth]{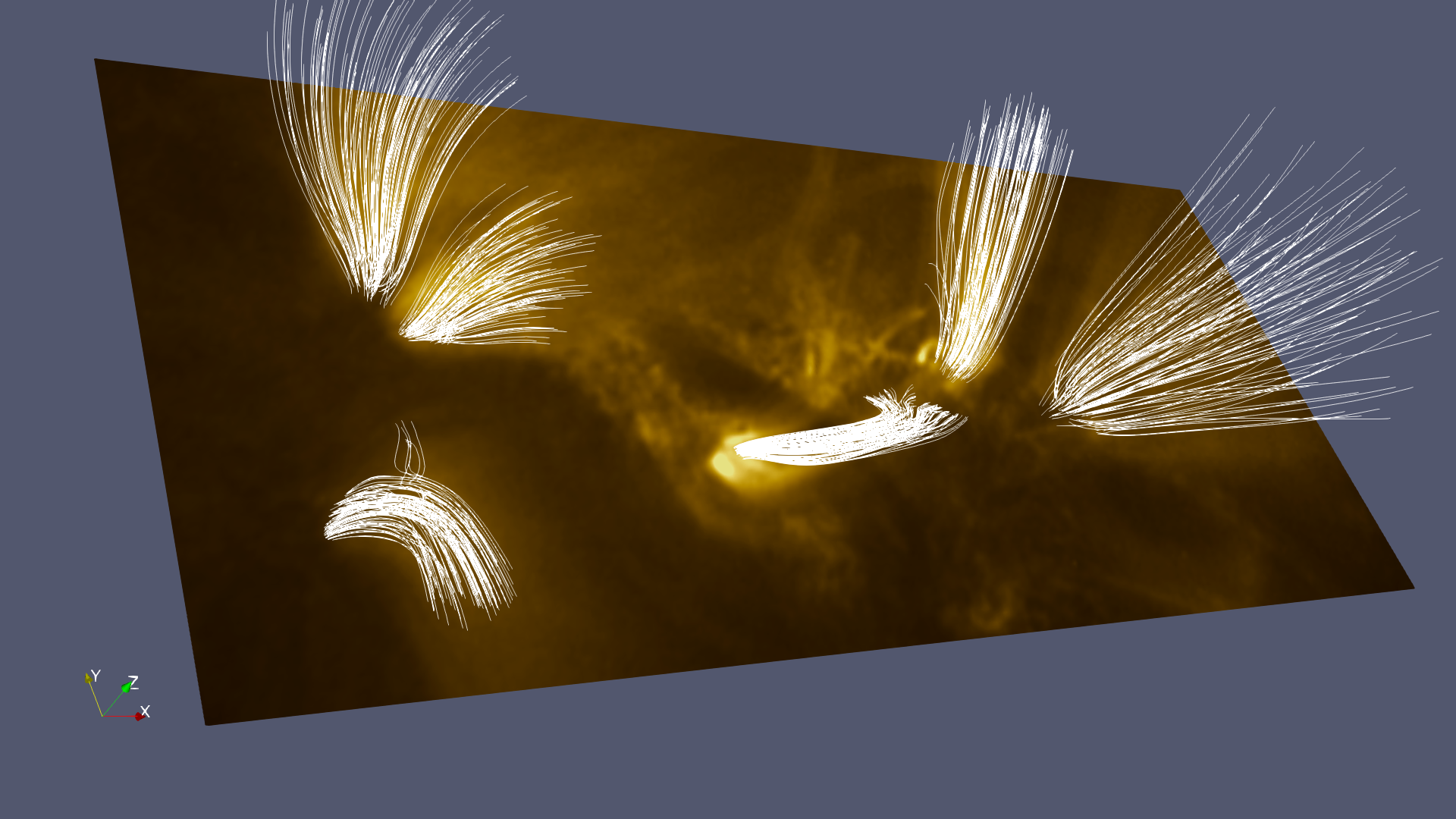}
    \includegraphics[width=0.48\textwidth]{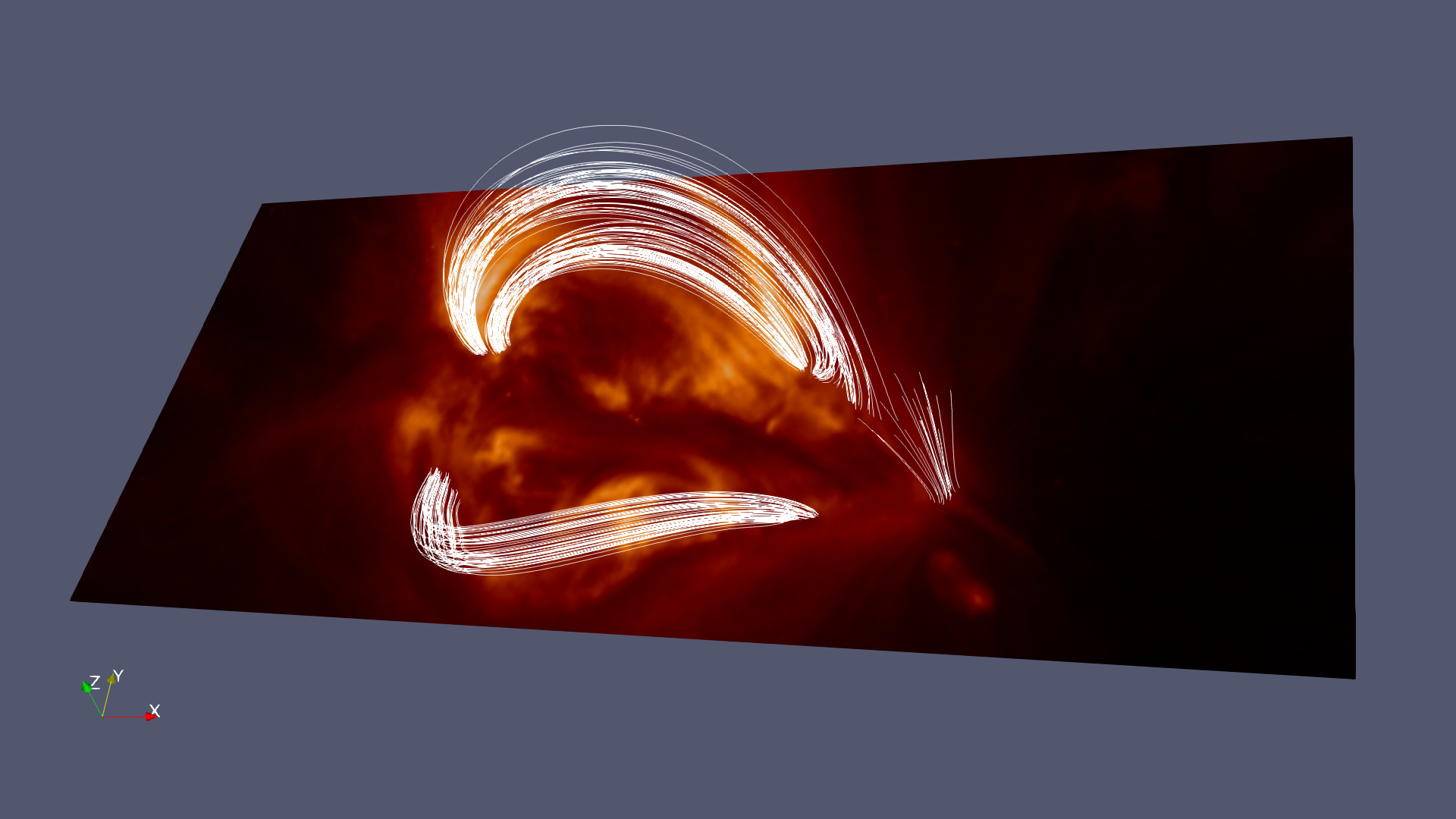}
    \includegraphics[width=0.48\textwidth]{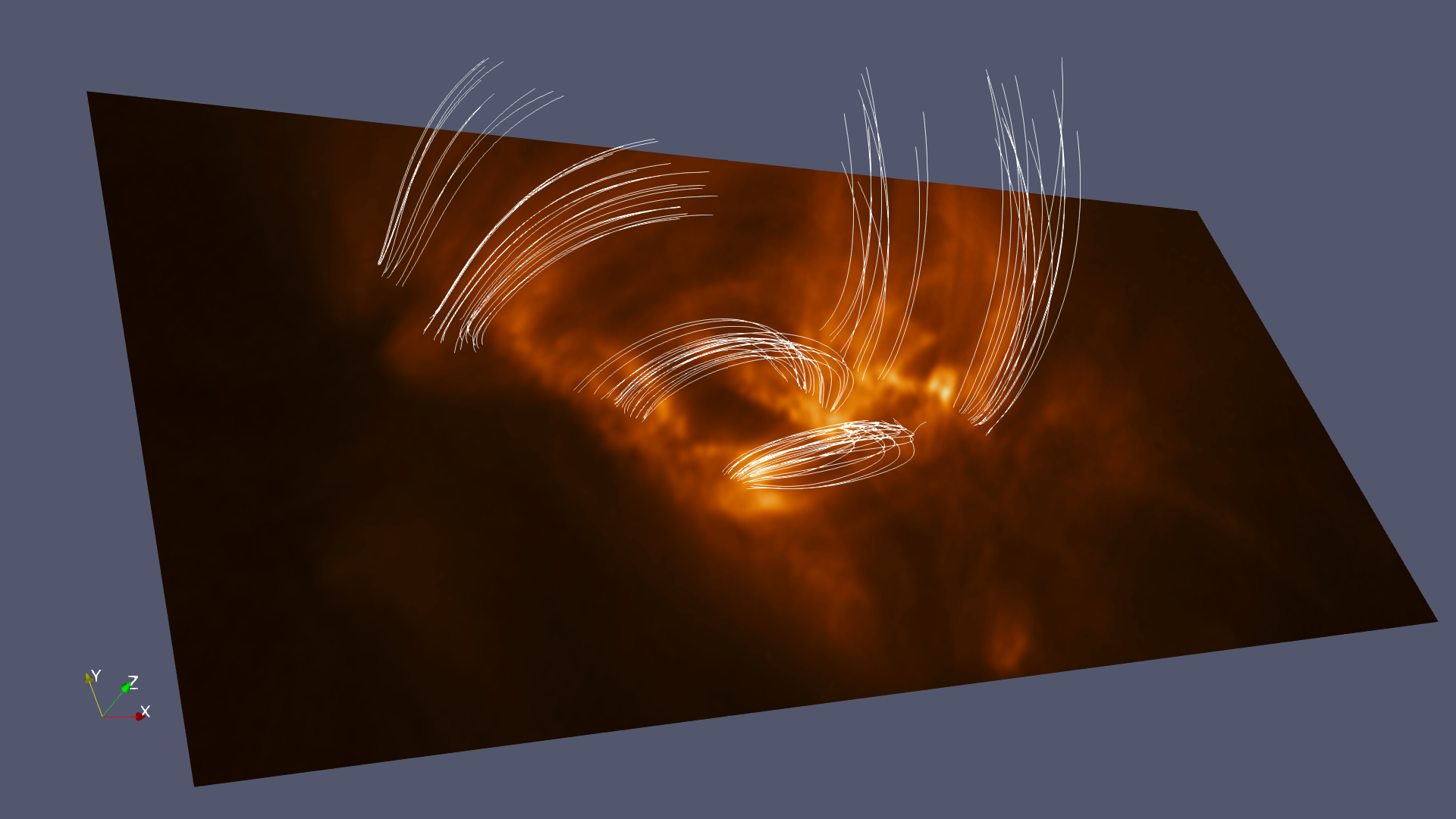}

    \caption{Comparison of extrapolated field lines to AIA observations from the AIA 171\AA\ (top row) and 193\AA\ (bottom row) channel on 2020 April 3 (left) and April 7 (right). The streamlines show that there is a good match between magnetic extrapolation used in this paper and the EUV observations.}
    \label{fig:extrap_euv}
\end{figure*}

\subsection{F10.7~cm Radio Flux Observation}

Radio observations were made by the JVLA on 2020 April 3 and 7 in the C array configuration~\citep{Perley2011Aug}. The observations were made between 14:40-21:20 UT on April 3 and between 15:00-22:05~UT on April 7 in the 2-4 GHz frequency band. The frequency band was subdivided into 16 sub-bands, or spectral windows, each with 128 MHz bandwidth. They were each observed with 64 frequency channels of 2 MHz. An integration time of 2s was used throughout. The high time and frequency resolution enabled radio frequency interference to be identified and excised from a given spectral window. 3C48 was used as the flux and bandpass calibrator and J0059+006 was used as the gain calibrator on both days. The observations were made in full polarization mode, allowing maps in total intensity (Stokes I) and circularly polarized intensity (Stokes V) to be formed. 

Since the field of view of the JVLA is $\sim 15$ arcmin, a mosaicking imaging strategy was employed to map the full disk of the Sun; i.e., 19 overlapping fields (Nyquist sampling) were used to provide full-disk coverage. For the present work, we focus only on those pointings in which AR 12759 was present and therefore formed maps using only three pointings on each date. Understanding observations from the JVLA and EIS are not cotemporal, we investigated the temporal evolution of the active region using AIA observations. On April 3, the active region was stable; and on April 7, the active region shows a minor filament activation between its leading and following polarity. This filament has no effect on our analysis. Interferometric instruments such as the JVLA serve as high-pass filters, resolving out emission on large angular scales. For the C array configuration, the background solar disk was effectively resolved out. However, if the total flux from the Sun is known, the background disk can be restored. We did so using a modified version of the feathering technique (Cotton 2015) and the daily observed F10.7 flux densities from the Dominion Radio Astrophysical Observatory \footnote{https://www.spaceweather.gc.ca/forecast-prevision/solar-solaire/solarflux/sx-4a-en.php}. 

The resulting maps (shown in Figure~\ref{fig:result}) provide radio images of AR 12759 with an angular resolution of approximately $9"$. These were converted to units of Kelvin (brightness temperature) in both Stokes I and V. On April 3 the maximum brightness temperature in the active region in the Stokes I map was $T_B=2.72\times 10^5$ K and on April 7 it was $T_B=3.74 \times 10^5$ K. The distribution of brightness temperatures on the background disk peaks at $T_B=3.7 \times 10^4$ K. The Stokes V maps on each day clearly show the bipolar nature of AR 12759. The degree of circular polarization of the active region emission, defined as $\rho_c=V/I$, is low on both days: ranging between -6.6\% and +10.2\% on April 3 and between -4\% and +5.8\% on April 7.

\subsection{Magnetic Extrapolation and Loop Connectivity}

The contribution of thermal gyroresonance emission to F10.7~cm originates in active regions from a thin layer where the emitted frequency (2.8~GHz)  is resonant with a low harmonic of the electron gyrofrequency. For coronal conditions, this occurs at the 2nd harmonic layer ($s=2$) or, more typically, at the 3rd harmonic ($s=3$) layer (e.g., \citealp{White1997Aug}), corresponding to $B=500$ G and 333~G, respectively. To locate the F10.7~cm gyroresonance emission sites, we modelled the coronal magnetic field of AR~12759 with linear fields extrapolated from photospheric magnetograms using the method of \citet{alissandrakis1981}~(Figure~\ref{fig:extrap_euv}). This method uses Fourier transforms to find the coronal magnetic field in a volume that satisfies the boundary conditions, including an observed magnetogram at the lower boundary. Many studies have modelled linear force-free magnetic fields using this methodology (e.g., \citealp{Green2002SoPh..208...43G,Yardley2022ApJ...937...57Y,James2022A&A...665A..37J}, to name just a few). One limitation of this method is that the maximum value of the force-free parameter, $\alpha$, that may be used is constrained by the spatial dimensions of the volume~\citep{Pevtsov1995}. If $\alpha$ is set too large for the chosen volume, the resulting field will be unphysical, with infinite energy. The boundary magnetograms are taken by the SDO/HMI, and are specifically from the SHARP data series (Spaceweather HMI Active Region Patch; \citealp{Bobra2014}). This data series provides information about the three-dimensional magnetic field vector in cutouts of the solar surface that contain one active region or more in a cylindrical equal-area (CEA) projection. Each pixel in the CEA projection represents an angular width of 0.03~degrees, or approximately 0.36~Mm.

We used an iterative method to determine the value of $\alpha$. We limited the field-of-view of the magnetogram used in this part of the procedure to include the full extent of strong magnetic field associated with AR 12759 whilst excluding as much quiet Sun noise as possible, and furthermore we only examined pixels where the horizontal field strength is greater than 200~G. We found the best values of $\alpha$ were $0.06~\mathrm{CEA-deg}^{-1}$ on 3rd April, and $-0.2 ~\mathrm{CEA-deg}^{-1}$ on April 7, which are both less than the maximum $\alpha$ that would still give real solutions in a volume based on the full SHARP magnetogram size. Therefore, we finally modelled the coronal magnetic field of AR 12759 by extrapolating the radial field component of the full HMI SHARP magnetograms taken at 13:36~UT on April 3 2020 with $\alpha  = 0.06~\mathrm{CEA-deg}^{-1}$ and 16:00~UT on April 7 2020 with $\alpha = -0.2 ~\mathrm{CEA-deg}^{-1}$ (Figure~\ref{fig:extrap_euv}).

We find good correspondence between selected field lines in the extrapolated fields and coronal loops observed in the 
EUV channels of SDO/AIA at the same times as the boundary magnetograms used in the extrapolations were taken (example of 171~\AA\ and 193~\AA\ are shown in Figure~\ref{fig:extrap_euv}), confirming that the linear magnetic fields represent the structure of AR 12759 at the selected times.

\begin{figure*}[!h]
    \centering    \includegraphics[width=0.9\textwidth]{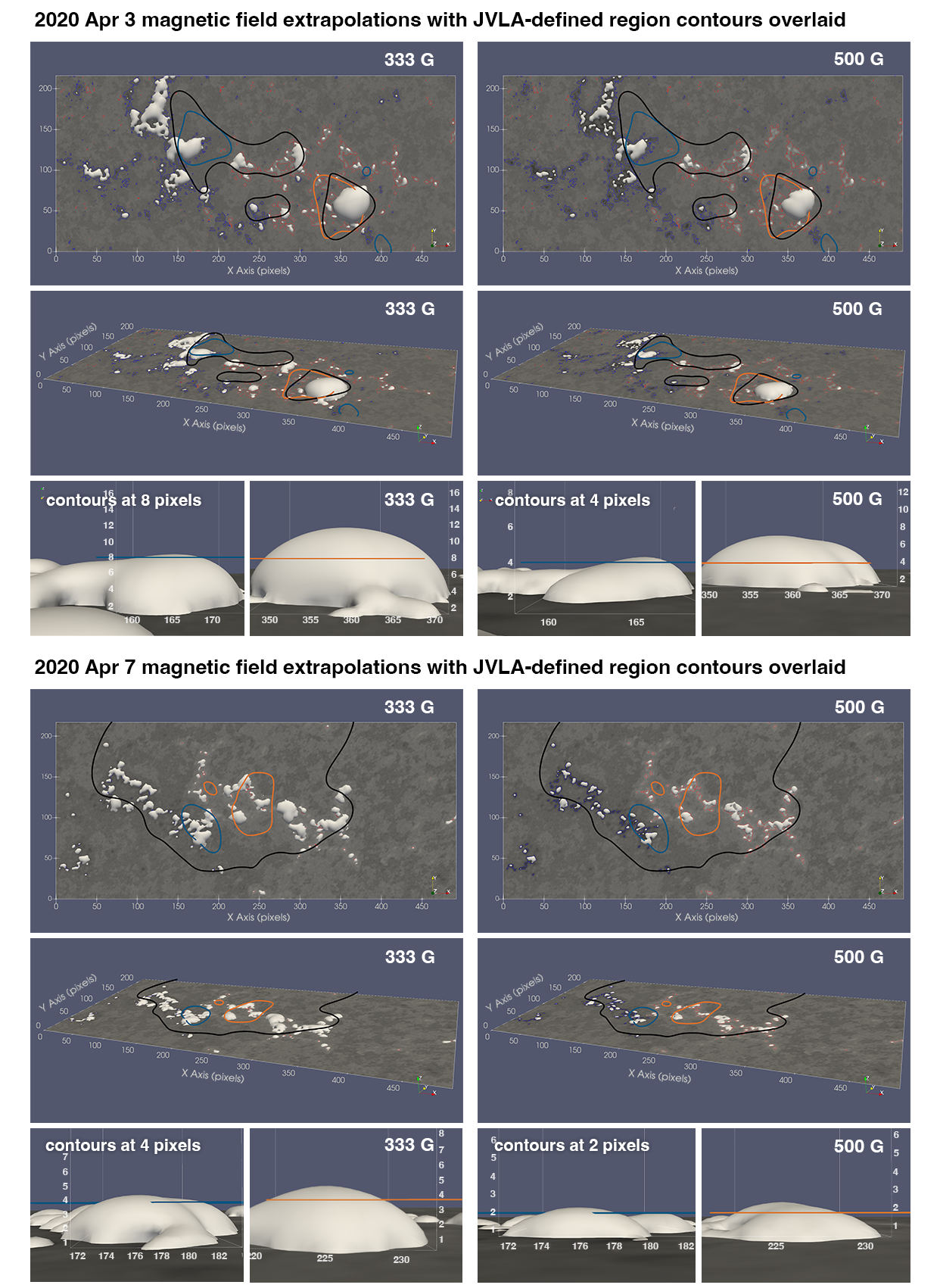}

    \caption{Comparisons between 333~G and 500~G isosurfaces (beige domes) in the magnetic field extrapolations to Stokes V emission measured by JVLA on 2020 April 3 (top) and April 7 (bottom). Positive and negative radial photospheric magnetic flux are contoured in red and blue on the HMI map, respectively. Thick black, blue and orange contours correspond to the regions defined using strong JVLA F10.7~cm emissions in Figure~\ref{fig:result}, with black: Stokes I $>80,000~$K; blue: Stokes~V $<-10,000~$K and orange: Stokes~V $>10,000~$K. On April 3, the 333 G domes around the JVLA emission reach a height of about 8 pixels (2.90 Mm) and the leading iso-gauss surface has a much higher height than the following surface, rendering gyroresonance emission. The AR is too weak on Apr 7 to render emission into groresonance.
}
    \label{fig:extrap_dome}
\end{figure*}

The 333 G and 500~G isosurfaces in the coronal magnetic field models of the active region on April 3 and April 7 are shown in Figure~\ref{fig:extrap_dome}. It can be seen that on April 3, the extrapolated iso-gauss surface of the leading polarity of AR~12759 reaches the greatest height compared to other regions/set of observations. We conclude that the emission from the lead spot on Apr 3 is consistent with gyroresonance emission. We placed the JVLA emission maps at different heights in the extrapolation volume to find the height where there was the closest match between the spatial extent of the strong Stokes~V emission and the isosurfaces of magnetic field strength. On April 3, there is a good match between locations of $\pm$10,000~K Stokes~V emission and 500~G and 333~G radial field strengths. We find that the best spatial match between the Stokes V emission is with the 333 G isogauss surface ($s=3$) at a height of 2.9 Mm.

\begin{figure*}[ht!]
    \centering
    \includegraphics[width=\textwidth]{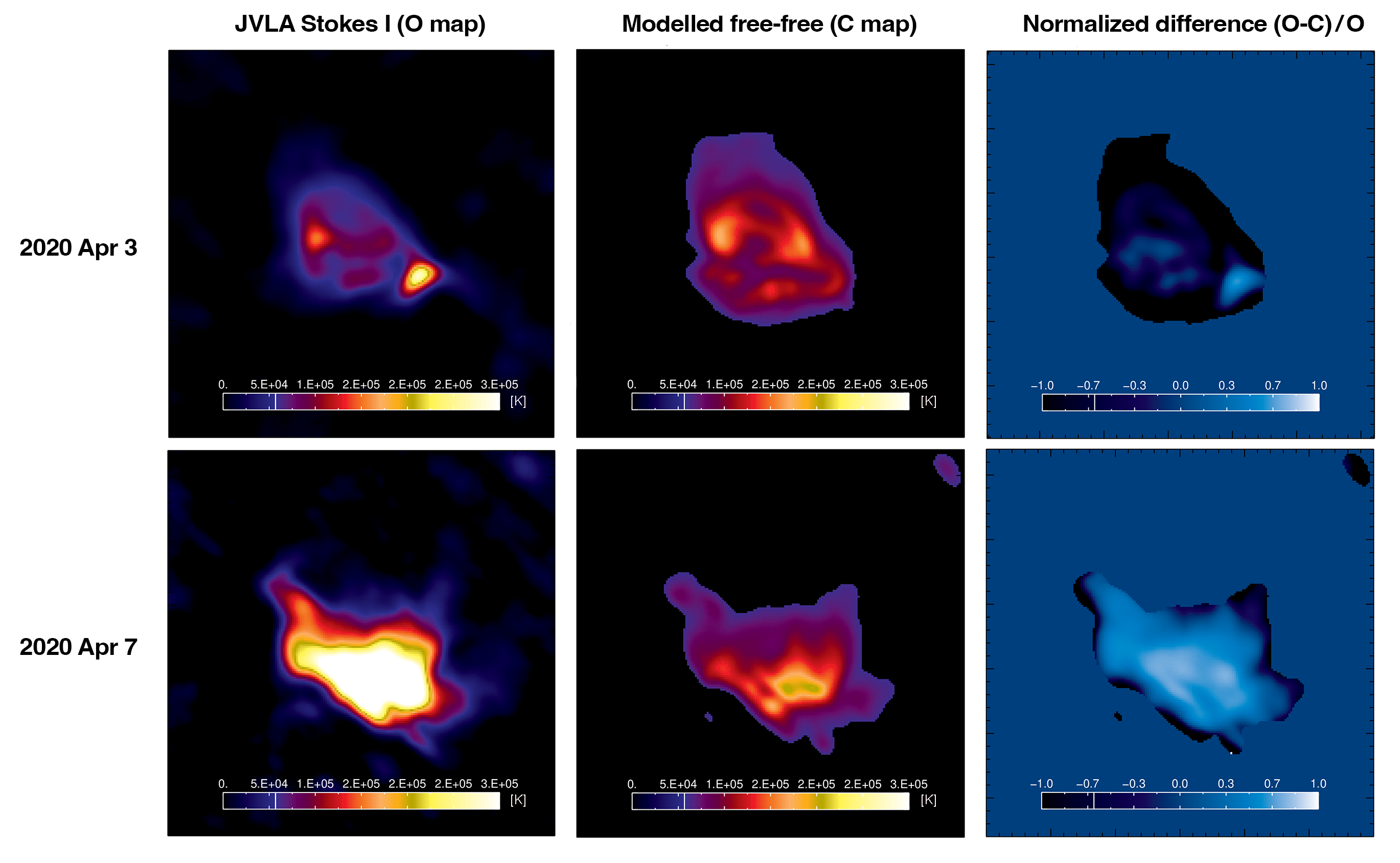}
    \caption{Top to bottom: Observations taken by JVLA/EIS on 2020 April 3rd and 7th. Left to right: JVLA Stokes I observation (O map), modelled F10.7~cm free-free emission calculated using AIADEM (C map)~\citep{Schonfeld2015Jul}, and the normalized difference, (O-C)/O map, used to estimate the location of F10.7~cm gyroresonance emission. The modelled free-free map was convolved using an elliptical Gaussian with the dimensions and position angle of the JVLA clean beam to mimic the JVLA observations. On April 3, it can be seen that the estimated gyroresonance region (normalized difference) is associated with the leading sunspot, with highest magnetic flux density; On April 7, the active region is dispersed and close to the limb, with possible influence of filament activation. Our analysis is therefore only based on the spot identifiable in the April 3 normalized difference map. Except the estimated gyroresonance region on April 3, most pixels in the normalized difference maps are within $\pm$ 0.5, consistent with the results in \citet{Schonfeld2015Jul}.} 
    \label{fig:extension}
\end{figure*}


\section{Results}\label{results}

\begin{table*}[]
\begin{tabular}{llccl}
\hline

             & Contour regions        & FIP Bias & Magnetic Flux Density & Note               \\ \hline
             \multicolumn{5}{l}{\textbf{Stokes~I and Stokes~V contours}}               \\ \hline

2020 April 3 & Total Intensity (Stokes~I)              & 2.7 &  19.2~G  &                    \\
             & Positive Stokes~V & 2.6     & 80.2~G & Leading polarity; Orange contour   \\
             & Negative Stokes~V & 3.9     & 73.7~G & Following polarity; Blue contour \\ \hline
2020 April 7 & Total Intensity (Stokes~I)              & 1.6 & 39.1~G    &                    \\
             & Positive Stokes~V & 1.7  & 67.3~G  & Leading polarity; Orange contour   \\
             & Negative Stokes~V & 1.9  & 42.1~G  & Following polarity; Blue contour \\ \hline
            \multicolumn{5}{l}{\textbf{Estimated gyroresonance contours ((Stokes~I-modelled free-free)/Stokes~I)}}              \\ \hline
2020 April 3
             & Gyroresonance  & 3.0  &  44.9~G &    Leading polarity; Red dashed contour\\\hline

\end{tabular}
\caption{Table containing the calculated FIP bias values and magnetic flux density associated with the two region defining methods. Top two rows: Regions associated with 1) strong Stokes~I, 2) positive and 3) negative Stokes~V profiles of the F10.7~cm emission, taken on both April 3 and 7. Bottom row: Region associated with 1) normalized difference (estimated gyroresonance emission region) using (Stokes~I-modelled free-free)/Stoke~I on April 3.}\label{table:result}
\end{table*}

Figure~\ref{fig:result} shows AR~12759 observations obtained on 2020 April 3 and 7. AIA 193~\AA\ images were used as context, followed by HMI magnetogram, FIP bias, F10.7~cm Stokes~V~(proxy for gyroresonance) and Stokes~I~(total intensity) maps. From the AIA~193~\AA\ full disk images shown in Figure~\ref{fig:result}, we can see that our observations were made when there was minimal solar activity, with AR~12759 the only active region on disk at the time. Although this active region was small, strong polarised emission can still be observed in the Stokes~V map, and the total intensity map traces out the overall morphology of the active region nicely. Using JVLA observations, the AR can be dissected into three parts. These three parts are:
\begin{enumerate}
    \item Leading polarity with strong positive Stokes~V emission, with brightness temperature $\mathrm{>10,000~K}$;
    \item Following polarity with strong negative Stokes~V, with brightness temperature $\mathrm{<-10,000~K}$;
    \item The overall active region indicated by the Stokes~I map with a brightness temperature $\mathrm{>80,000~K}$, subtracted by the strong Stokes~V regions defined above. This indicates that the region is dominated by free-free~(bremsstrahlung) emission.
\end{enumerate}
Three regions were then defined to investigate the relationship between coronal abundances and radio F10.7 flux. These values were chosen to include most of the strongest emitting regions (Figure~\ref{fig:result}). We then averaged the EIS observed intensities in each of these three regions, and calculated the spatially averaged composition value, with results listed in Table~\ref{table:result}. The spatially averaged FIP bias are assumed to have an error of 0.3. Figures~\ref{fig:extrap_euv} and \ref{fig:extrap_dome} show the magnetic field extrapolation of AR~12759. In Figure~\ref{fig:extrap_dome}, white contours indicating strong Stokes~V emission are plotted on top of the iso-gauss surfaces. We see good correlation between Stokes~V emission and areas with strong magnetic field strength. Although AR~12759 was small with weak magnetic field strength, distinct differences between the coronal abundances can be observed when the active region was stronger on 2020 April 3. On April 3, over the region with positive Stokes~V emission~(leading sunspot), FIP bias is around 2.6. A similar FIP bias value can be observed in the free-free emitting region~(FIP Bias = 2.7), with the highest FIP bias observed in the negative Stokes~V region~(FIP Bias = 3.9).

However, differences between FIP bias values associated with the three sub-regions were much smaller or non-existent on April 7. As shown in the HMI magnetogram in Figure~\ref{fig:result}, the active region was much weaker on April 7, with no distinct identifiable sunspot. Both of the polarities are much more dispersed on April 7. This seems to have a significant lowering effect on the overall coronal abundance, and FIP bias remains roughly the same over the three regions, with free-free~(FIP Bias = 1.6); positive Stokes~V~(FIP bias = 1.7); and negative Stokes~V~(FIP bias = 1.9).

\section{Discussion}\label{discussion}

So far, we have analysed the relationship between FIP bias and Stokes~I and Stokes~V of the F10.7~cm emission. In our first set of observations on April 3, clear differences in FIP bias can be observed between the Stokes~I, positive and negative Stokes~V regions. Significantly enhanced coronal abundances can be observed associated with the negative Stokes~V region (following polarity), whereas we see a much lower FIP bias associated with the positive Stokes~V (leading polarity) region. For the second set of observations taken on April 7, this differentiation between sub-regions completely vanishes. Although we defined contours using the same parameters across the two days, all three regions have roughly the same low FIP bias value of $\sim$1.7. This inconsistent behaviour is extremely interesting, and looking into Figure~\ref{fig:result} and \ref{fig:extrap_dome}, the most obvious difference across the two sets of observations is the magnetic field strength or magnetic flux density of each region. From the HMI magnetogram shown in Figure~\ref{obs}, on April 3, the magnetic fields are more closely bound together, whereas by April 7, the magnetic fields are much more dispersed. By comparing FIP bias to the Stokes~I \& V profile of the F10.7 radio flux, these results confirm that Stokes~V F10.7~cm radio emission comes from the highest magnetic magnetic field strength or magnetic flux density areas in the AR. However, not all Stokes~V emission may come from gyroresonance emission. In the next section, we try to isolate gyroresonance regions.

\subsection{Region Associated with F10.7~cm Gyroresonance Emission} \label{gyro_region}

Since the Stokes~V profile contains all of the polarised signal, not only gyroresonance emission, our Stokes~V map is inevitably mixed with polarised free-free emission. In order to check if F10.7~cm gyroresonance emission also plays a role in contributing to different FIP biases, we utilise the relationship to relate coronal bremsstrahlung (free-free) emission and DEM,
\begin{dmath}\label{equ:density}
    f_{\nu} = 9.78\times10^{-3}\frac{2k_B}{c^2}\left(1+4\frac{n_{\mathrm{He}}}{n_{\mathrm{H}}}\right)
    \times\iint T^{-0.5}~\mathrm{DEM}(T)\mathrm{G}(T)~dT d\Omega ,
\end{dmath}
where $k_B = \mathrm{1.38\times 10^{-16}~g~cm^2~s^{-2}~K^{-1}}$ is the Boltzmann's constant, $c = \mathrm{3\times10^{10}~cm~s^{-1}}$ is the speed of light, $n_{\mathrm{He}}/n_{\mathrm{H}} = \mathrm{0.085}$ is the density ratio of helium to hydrogen in the corona, $T(K)$ is the temperature, $\mathrm{G}(T)~=~\mathrm{24.5~+~}\ln(T/\nu)$ is the Gaunt factor, where $\nu$ is in Hz, and $d\Omega$ is the solid angle of the source~\citep{Dulk1985Sep}. Equation~\ref{equ:density} shows that the JVLA free-free emission can be estimated using a DEM. Therefore, by calculating the normalize difference ((observed Stokes~I - modelled free-free)/observed Stokes~I), the remaining signal should indicate the locations of strong gyroresonance emissions.

To model F10.7~cm free-free emission, we used the regularized inversion technique described in \citet{Hannah2012Mar} to derive a DEM using the AIA instrument (AIADEM) for both dates. The derived DEM is then inserted into Equation~\ref{equ:density} to generate the modelled free-free map. To achieve results similar to the JVLA observation, we convolved the calculated free-free map using an elliptical Gaussian with the dimensions and position angle of the JVLA clean beam. The results are shown in Figure~\ref{fig:extension}, where we have plotted the JVLA Stokes I (O map), modelled free-free emission (C map), and the normalised difference map (O-C)/O.  On April 3, while AR~12759 was still relatively intact, the modelled free-free map shows very good agreement with the JVLA Stokes~I map. However, on April 7, the active region had dispersed, no distinct spot can be identified in the normalized difference map. Also, given that the 333 and 500~G iso-gauss surfaces are very likely to be within the optically thick layer (Figure~\ref{fig:extrap_dome}), gyroresonance emission makes no significant contribution to the observed radio emission. The under-estimation of the free-free model could be due to filament-activation related activities on April 7. Except the regions mentioned above, most pixels in the normalized difference maps are within $\pm$ 0.5, consistent with the results in \citet{Schonfeld2015Jul}.

To investigate the relationship between F10.7~cm gyroresonance emissions and FIP bias, we focus on the April 3 data. It can be seen from the difference map (Figure~\ref{fig:extension}; 3rd column) and the HMI magnetogram (Figure~\ref{fig:result}; red dashed contour), the observed to predicted maps deviate in the area associated with the leading polarity. As a confirmation, we also used the EIS MCMCDEM technique to model the April 3 spatially averaged free-free emissions at the 3 regions defined and illustrated in Section~\ref{results} and Figure~\ref{fig:result} respectively. Using the different DEM calculation method, the modelled free-free emission behaves similarly to the free-free emission calculated by AIADEM. The blue contour in Figure~\ref{fig:result} (following polarity) has the highest modelled free-free emission, followed by the black contours (region between polarities), and finally the orange contour (leading polarity). This is consistent with the AIADEM method, where a large deviation only exists over the leading polarity. From the magnetic extrapolation, the iso-gauss surface associated with the leading spot reaches a height of about 8 pixels (2.90 Mm). It is likely that part of this surface is located at the optically thin region, further suggesting that this is the location of the F10.7~cm gyroresonance source. Therefore, we repeated our composition calculation on this region, and the results are shown in Table~\ref{table:result}.

Interestingly, using this second region defining method using the estimated gyroresonance location, the FIP bias remains roughly the same, maintaining a value of ~3.0. These unchanged FIP bias values, yet again, gives us a hint of the change in F10.7--coronal abundances correlation during different levels of solar activity stated in \citet{Brooks2017Aug}.

\subsection{Interpretation} \label{interpretation}

Overall, we calculated the FIP bias using two different region defining methods. One taken straight from the JVLA Stokes~I and Stokes~V map, the other one using the gyroresonance emission region estimated using DEM calculated using the AIA instrument. From the first method, the leading region has a slightly enhanced FIP bias value, at 2.6, whereas the following region shows significantly enhanced coronal abundances, at FIP bias = 3.9. As we move on to the second region defining method, using a region associated with gyroresonance emission, the leading spot shows a FIP bias of around 3.0, a value that is still a lot lower than the FIP bias found in the more dispersed negative-Stokes V area. We believe that the combination of these two methods tells  the same story, magnetic field strength plays a crucial role in the variation of coronal abundances. On April 3, the active region still contained sunspots, and the overall FIP bias value is higher when compared to observations taken on April 7. However, as we zoom into the small sub-regions, different magnetic concentrations contribute differently to the FIP bias observed. In both of the region-defining methods, the leading polarity has always been associated with a stronger, more concentrated magnetic field. Under this configuration, the magnetic fields associated with this emission inhibited convection. The consequent cooler temperatures lead to a lower ionization rate, thus a slightly lower FIP bias~\citep{Baker2021Jan, Mihailescu2022May}. In contrast, in the following spot, roughly the same magnetic flux is spread out into a larger area. The higher temperatures lead to a higher ionization rate of the low-FIP elements. Our result shows similar behaviour to the sunspots investigated in~\citet{Mihailescu2022May}, who also found a slightly lower FIP bias value in the leading polarity sunspot region with higher magnetic flux density.

This result can be translated into a bigger picture, informing us on the relationship between Sun-as-a-star FIP bias and F10.7 flux. According to the ponderomotive force model of fractionation developed by \citet{Laming2015Sep}, nanoflares caused by the reconnection of braided magnetic field in the corona trigger Alfv\'{e}n waves that travel down to the field line's footpoint in the chromosphere. These Alfv\'{e}n waves are being repeatedly refracted and reflected in the strong density gradient of the chromosphere, initiating the ponderomotive force which brings ions to the corona, contributing to what we quantify as FIP bias~\citep{Laming2015Sep, Laming2021Mar}. In this context, when the solar activity is low, overall coronal abundances behave similarly to AR~12759 on April 7. Magnetic flux is low in the activity belt, and the lower nanoflare activity and consequently lower resonant Alfven-wave activity in coronal loops results in a lower Sun-as-a-star FIP bias during solar minimum.Then, as solar activity ramps up, more regions of strong field and nanoflares contribute to more Alfv\'{e}n waves being created and reflected along closed magnetic loops. FIP bias slowly goes up with the solar activity. As F10.7~cm radio flux is a proxy for solar activity, we expect Sun-as-a-star FIP bias to correlate extremely well with F10.7~cm measurements under the above scenarios. However, during peak solar activity, more and more gyroresonance emissions start to mix into the F10.7 index. As a whole, although we have higher F10.7 flux, FIP bias values stop changing, resulting in the change in correlation we see in \citet{Brooks2017Aug}.


\section{Conclusions}\label{conclusion}

In this paper, we present observations of a small active region, AR~12759, using Hinode/EIS, JVLA and SDO/AIA. This active region was in its decay phase, initially containing a leading sunspot and trailing pores before decaying to only dispersed magnetic flux with no spots on 2020 April 7. There are significant differences between the magnetic field strengths (flux densities) observed on 2020 April 3 and 7. We employed two region defining methods to investigate the relationship between FIP bias and different emission mechanisms of the F10.7 radio flux. One defined using the Stokes~I and Stokes~V maps, the other using the estimated gyroresonance region isolated with the help of both Stokes~I map, and AIADEM. Combining the results from the two methods, we find that the following-polarity region carries a significantly enhanced Si/S coronal abundance. In contrast, in the leading polarity, no matter how we alter the region defining method, the FIP bias enhancement seems to be weaker, maintaining a value slightly higher than the quiet Sun.

This analysis is consistent with the findings in \citet{Brooks2017Aug}. Under low-medium solar activity (or in other words, when there is no/low gyroresonance emission), magnetic flux density plays an important role in varying elemental fractionation. At the start of the solar cycle, active regions (their sunspots) are smaller than the ones that emerge later into the cycle~\citep{Watson2011, Valio2020}. Sun-as-a-star FIP bias rises with the appearance of each new active region and the increased heating rate within. As each active region evolves, their decay (dispersion to a plage-like magnetic flux density) further increases the overall FIP bias, as we found when comparing the spot-containing leading and the plage-like following-polarity areas in AR 12759 on April 3. At the end, as we found on April 7, during the late decay phase of an AR, FIP bias decreases. Under such low-activity, low-gyroresonance conditions, FIP bias and F10.7~cm emission show a good correlation.

However, as activity rises towards the solar maximum, the maximum sunspot area increases, spots have higher field strength and become cooler \citep{Watson2011, Valio2020}. This was, in particular, confirmed being the case for the 2009--2014 period by \citet{Rezaei2015}. We suggest that under high solar activity conditions with rising sunspot area and field strength, the contribution of gyroresonance emission to the F10.7~cm emission will likely increase. As with the second method we found that while coronal abundances maintain roughly the same level in regions of high magnetic flux density, the gyroresonance radio flux from these spotted areas is not lower, but significantly higher than from plage regions. So we postulate that the distinctly different tendencies of radio gyroresonance emission and coronal abundances (FIP bias) over strong magnetic field concentrations (sunspots) is the likely cause of the saturation of Sun-as-a-star coronal abundances around solar maximum. Amid this FIP bias saturation, the F10.7 index, with its combined contribution from both free-free (bremsstrahlung) and gyroresonance components, remains well correlated with the sunspot number and other magnetic field proxies, creating this nonlinear correlation in \citet{Brooks2017Aug}.

Our observations provide a glimpse into the reason behind the nonlinear relationship during solar maximum. However, the fact that our JVLA--EIS joint observations were made during solar minimum, on a small active region relatively close to the solar limb has limited our ability to further investigate the relationship between magnetic field behaviour and FIP bias. Taking several observations of different active regions when the Sun is more active would be important to confirm our finding. Ideally, a statistical sample of EIS and JVLA maps could be built up from solar min to solar max to fully understand the correlation and scaling between F10.7 flux and coronal abundances. Also, although we have gone to great lengths to minimise the effects that arise from differences in the line of sight, observations of Sun-centred active regions could further reduce the alignment uncertainty. Since F10.7 radio flux comes from a wide range of solar altitudes, adding observations using the Solar Orbiter, in particular, Spectral Imaging of the Coronal Environment (SPICE), can add another layer of analysis, and further constrain our results from the corona to the chromosphere. The upcoming Solar-C EUV High-throughput Spectroscopic Telescope (EUVST) and its wide range of cotemporal temperature coverage can also contribute massively by observing different layers of our Sun’s atmosphere simultaneously. It is worth noting that in this paper, we have focused on observations of one active region. It is important not to dismiss that during solar maximum, we have many more on disk coronal holes and intense flares. These coronal holes and flares also play a role in lowering the Sun-as-a-star FIP bias during solar maximum. Although our finding shows great consistency with previous studies, more observations on coronal holes/flares are required to precisely disentangle the relationship between F10.7~cm and Sun-as-a-star FIP bias. 

Apart from solar composition, our result could be extended to the context of stellar coronal composition. Low activity stars like our Sun have coronae that are dominated by the FIP effect (a more enhanced low-FIP composition). This result highlights the importance of magnetic magnetic field strength or magnetic flux density and the F10.7~cm emission when linking coronal composition to the different spectral types of stars. In addition to the stellar coronal composition investigation done in \citet{Wood2010Jun, Seli2022A&A...659A...3S}, a full-cycle observation should also be considered to fully understand stellar composition variability.

\section{Appendix}

\begin{figure*}[h!]
    \centering
    \includegraphics[width=\textwidth]{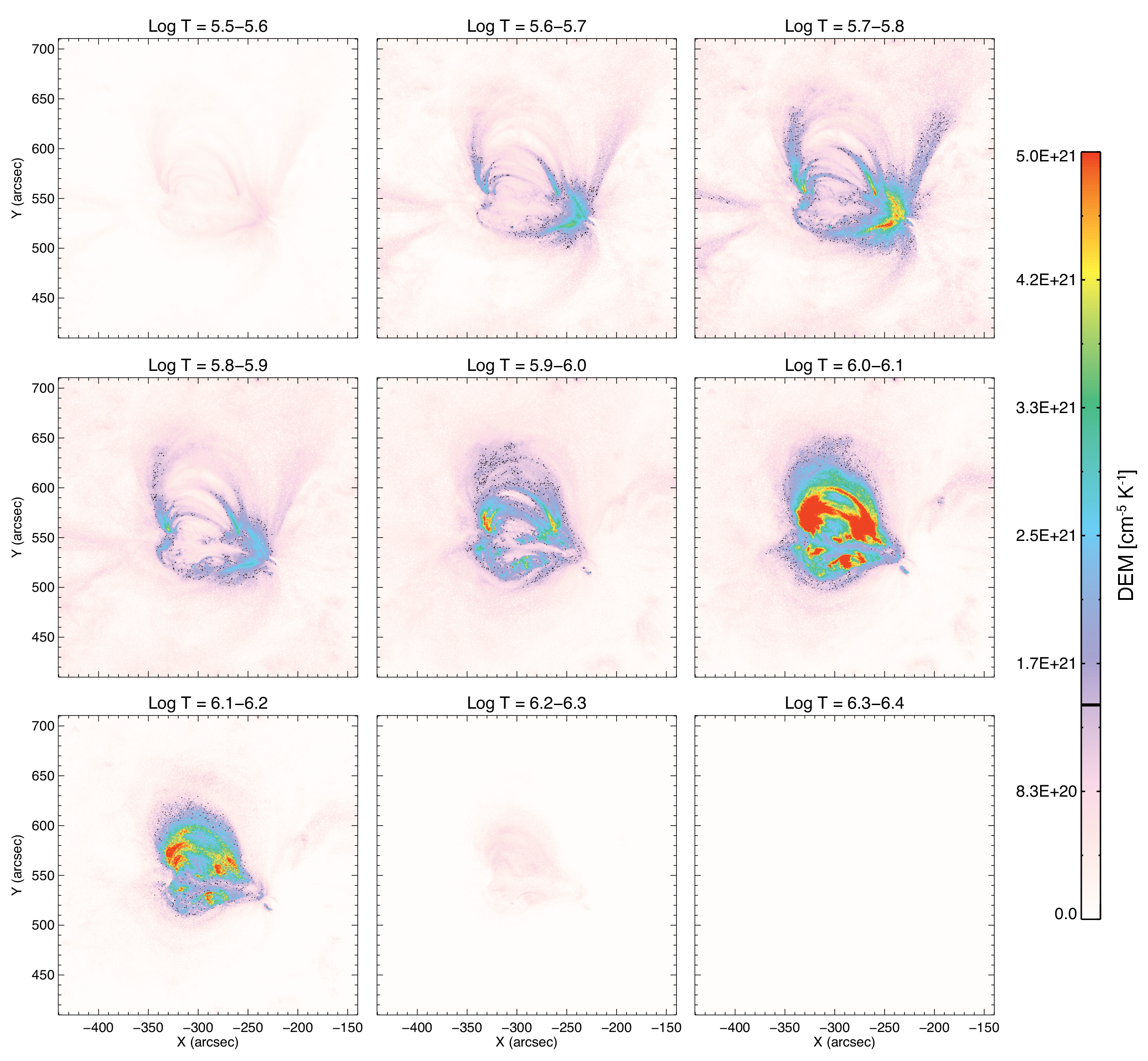}
    \caption{DEM on April 3 calculated using AIA passbands~\citep{Hannah2012Mar}. The DEM shows that the simple active region has a relatively low temperature, with a majority of emission from $\mathrm{log T = 6.0-6.1}$.}
    \label{fig:aiadem}
\end{figure*}

In this section, we show the AIA DEM used to model the F10.7~cm free-free emission on 2020 Apr 3. The DEM in Figure~\ref{fig:aiadem} shows that the core loops of the active region has a temperature of logT = 6.0--6.1.

\acknowledgements{
We thank the reviewer for the very constructive comments. A.S.H.T. thanks the STFC for support via funding given in his PHD studentship. A.W.J. is supported by a European Space Agency (ESA) Research Fellowship. D.M.L. is grateful to the Science Technology and Facilities Council for the award of an Ernest Rutherford Fellowship (ST/R003246/1). The work of D.H.B. was performed under contract to the Naval Research Laboratory and was funded by the NASA Hinode program. D.B. is funded under STFC consolidated grant number ST/S000240/1 and L.v.D.G. is partially funded under the same grant. L.v.D.G. acknowledges the Hungarian National Research, Development and Innovation Office grant OTKA K-113117. G.V. acknowledges the support from the European Union's Horizon 2020 research and innovation programme under grant agreement No 824135 and of the STFC grant number ST/T000317/1. Hinode is a Japanese mission developed and launched by ISAS/JAXA, with NAOJ as domestic partner and NASA and STFC (UK) as international partners. It is operated by these agencies in co-operation with ESA and NSC (Norway). AIA data courtesy of NASA/SDO and the AIA, EVE, and HMI science teams. CHIANTI is a collaborative project involving George Mason University, the University of Michigan (USA) and the University of Cambridge (UK). The National Radio Astronomy Observatory is a facility of the National Science Foundation operated under cooperative
agreement by Associated Universities, Inc. I thank Dr Hamish Reid for a fruitful discussion after the FASR 2021 conference.
}

\bibliography{bib}{}
\bibliographystyle{aasjournal}



\end{document}